\documentclass[12pt]{article}
\usepackage{graphicx}
\usepackage{amssymb,amsmath}
\usepackage{mathtools}
\usepackage{textcomp}
\usepackage{multirow}
\usepackage[english]{babel}
\usepackage{color}
\usepackage{epstopdf}
\usepackage{caption}
\usepackage{subcaption}
\usepackage{breqn}
\usepackage{hyperref}
\usepackage[
top    = 4cm, bottom = 4cm, left   = 3.5cm, right  = 3.5cm]{geometry}

\begin{document}

\newcommand{\ket}[1]{\ensuremath{\left|#1\right\rangle}}
\newcommand{\bra}[1]{\ensuremath{\left\langle#1\right|}}

\begin{center}
{\large{\bf $(t,n)$ THRESHOLD $d$-LEVEL QUANTUM SECRET SHARING BASED ON QUANTUM FOURIER TRANSFORMATION}}\\
\bigskip
SARBANI ROY \footnote{ sarbani16roy@gmail.com}, SOURAV MUKHOPADHYAY \footnote{ msourav@gmail.com}\\
\bigskip
Department of Mathematics, Indian Institute of Technology Kharagpur, Kharagpur - 721302, India\\
\end{center}

\begin{abstract}
Quantum secret sharing (QSS) is an important branch of secure multiparty quantum computation. Several schemes for $(n, n)$ threshold QSS based on quantum Fourier transformation (QFT) have been proposed. Inspired by the flexibility of $(t, n)$ threshold schemes, Song {\it et al.} (Scientific Reports, 2017) have proposed a $(t, n)$ threshold QSS utilizing $QFT$. Later, Kao and Hwang (arXiv:1803.00216) have identified a loophole in the scheme but have not suggested any remedy. In this present study, we have proposed a $(t, n)$ threshold QSS scheme to share a $d$ dimensional classical secret. This scheme can be implemented using local operations (such as $QFT$, generalized Pauli operators and local measurement) and classical communication. Security of the proposed scheme is described against outsider and participants' eavesdropping.
\end{abstract}


\section{Introduction}
\label{intro}
Secret sharing (SS) is an important branch of cryptography. In 1979, Shamir~\cite{Shamir} proposed secret sharing scheme to share a secret among several participants in such a way that a set of authorized participants can reconstruct the secret. In general, classical secret sharing (CSS) can be classified in $(n,n)$ and $(t,n)$ threshold. In an $(n,n)$ threshold secret sharing scheme all $n$ participants should collaborate to reconstruct the secret, whereas in a $(t,n)$ threshold secret sharing scheme the cardinality of the authorized set is $t$. $(t, n)$ threshold SS has various applications in group authentication~\cite{harn}, threshold signature~\cite{bol,harn1}, group key agreement~\cite{liu1}, threshold encryption~\cite{des}, secure multiparty computation~\cite{patel}, etc.

With the development of quantum cryptography, quantum secret sharing (QSS) is receiving more and more interest. QSS which is a variation of CSS in quantum cryptography, uses quantum computation to share and reconstruct the secret. QSS schemes provide security depending on physical laws, whereas classical ones usually achieve security based on computational assumptions. Thus QSS schemes are more reliable. There are two types of QSS: sharing a classical secret and sharing a quantum secret. Hillery et al.~\cite{Hillery} for the first time proposed a QSS scheme that shares classical secret of a dealer Alice among two participants Bob and Charlie. In the scheme, Alice, Bob and Charlie share a three particle two dimensional GHZ state. They measure their own particle randomly in one of the two directions and announces their measurement basis publicly but not the measurement result. To gain Alice's measurement result, Bob and Charlie colludes with their measurement results. This allows them to establish a private key and hence share a secret. Cleave et al.~\cite{cleave} have proposed another QSS scheme that shares quantum secret using CSS code~\cite{cs,s}, a quantum error correcting code. After that a large number of schemes on QSS have been proposed such as circular QSSs~\cite{zhou,lin,zhu}, dynamic QSSs~\cite{hsu,wang}, threshold QSSs~\cite{KM,DFPR,BBK}, single particle QSSs~\cite{Schmid,tav,kar}, graph state QSSs~\cite{markham,keet,sar}, verifiable QSSs~\cite{yang1,yang2,sl} and QSSs based on error correcting codes~\cite{sk}, phase shift operations~\cite{qin,du,liu} and quantum search algorithms~\cite{hsu1}.

Yang {\it et al.}~\cite{yang3} have shown that the quantum Fourier transformation (QFT) can be used to propose an $(n,n)$ threshold QSS scheme in which the share of each participant is protected by true randomness. After having the share (or private shadow) from the dealer, the participants share an $n$ particle GHZ state in qudits. Each participant applies QFT followed by a unitary transformation (depending on private shadow) on his own particle. Then they measure their own particle and gets the dealer's secret by adding all the measurement results. But the share distribution procedure is not described in the scheme. They have only mentioned that secret is distributed in such a way that summation of the shares will return the secrets. To reduce the computational cost, Xiao and Gao~\cite{xiao} have proposed an $(n,n)$ threshold $d$ level QSS scheme based on local operation and classical communication (LOCC). Similar to \cite{yang3}, the dealer and the $n$ participants share a $(1+n)$ particle $d$ dimensional GHZ state. The dealer applies generalized Pauli $Z$ gate (depending on his secret) on his own particle. Then the dealer and all the participants apply $QFT$ on their own particles and measure those in the computational basis. Now, Dealer announces his measurement result and participants consider their measurement result as their private shadow. Finally, the participants recover the secret by adding their measurement results along with dealer's measurement result.

The above two QSS schemes based on $QFT$ are $(n, n)$ threshold. Whereas $(t, n)$ threshold QSS is more flexible than $(n, n)$ threshold QSS. Note that, to reconstruct the secret for a $(t, n)$ threshold QSS scheme, it requires the participation of any $t$ participants, whereas all $n$ participants have to contribute in case of $(n, n)$ threshold. Inspired by the flexibility of $(t, n)$ threshold QSS schemes, Song {\it et al.}~\cite{song} have proposed a $(t,n)$ threshold QSS scheme that shares $d$ dimensional classical secret using quantum $QFT$. Kao and Hwang~\cite{kao} have identified that the scheme fails to reconstruct the secret. But, they have not suggested any improvement of the scheme in~\cite{song} to mitigate this loophole.

In this current draft we have revisited the scheme proposed by Song {\it et al.}~\cite{song} and a loophole in that scheme. To mitigate the loophole, We have proposed a $(t, n)$ threshold $d$-level QSS scheme based on the idea of quantum Fourier transformation on a $d$-dimensional multi-particle entangled state. We have also verified the security of the proposed scheme against all possible outsider and participant's attack.

The rest of this paper is organized as follows. In section~\ref{sec:2} some correlative preliminaries are introduced. Section~\ref{sec:3} revisits the scheme proposed by Song {\it et al.}~\cite{song} and its loophole. Section~\ref{sec:4} explains the design of the method for the proposed scheme. Section~\ref{sec:5} proves the correctness. Section~\ref{sec:6} analyzes the security. Section~\ref{sec:7} compares our scheme to some of the existing schemes. Finally, in scetion~\ref{sec:8}, the conclusion of this paper is given.

\section{Preliminaries}
\label{sec:2}
Before describing the protocol, here in this section we have introduced some preliminary concepts necessary to describe the protocol, which includes the basic ideas of quantum Fourier transformation, generalized Pauli operator and the Shamir's $(t, n)$ threshold secret sharing scheme~\cite{Shamir}. Note that, for the rest of our discussion we have used `$+$' for addition modulo $d$ and `$\cdot$' for multiplication modulo $d$.

\subsection{Quantum Fourier Transformation}
The quantum Fourier transform (QFT) is a unitary transformation on a quantum system. For a basis $\{0, 1, \cdots, d-1\}$ in $d$ dimension, the $QFT$ is defined by

{\scriptsize
\begin{equation}
QFT\ket{y} = \frac{1}{\sqrt{d}}\sum_{x=0}^{d-1}\omega^{y\cdot x}\ket{x},
\end{equation}} where $\omega =e^{\frac{2 \pi i}{d}}$ and $y\in\{0, 1, \cdots d-1\}$. For the same basis, the inverse quantum Fourier transformation $(QFT^{-1})$ is defined by
{\scriptsize
\begin{equation}
QFT^{-1}\ket{x} = \frac{1}{\sqrt{d}}\sum_{y=0}^{d-1}\omega^{-x\cdot y}\ket{y},
\end{equation}} where $x\in\{0, 1, \cdots d-1\}$.

\subsection{Generalized Pauli Operator}
Let $H$ be a $d$ dimensional Hilbert space with a basis $\{\ket{0}, \ket{1}, \cdots, \ket{d-1}\}$. The generalized Pauli operator on $H$ is defined by
{\scriptsize
\begin{equation}
U_{\alpha, \beta} = \sum_{x=0}^{d-1}\omega^{\beta\cdot x}\ket{x+\alpha}\bra{x}.
\end{equation}}
In particular,
{\scriptsize
\begin{equation}
U_{l, 0}\ket{x}=\ket{x+l},~U_{0,l}\ket{x} = \omega^{l\cdot x}\ket{x},
\end{equation}}
for $l\in\{0, 1, \cdots d-1\}$, are called generalized Pauli $X$ gate and generalized Pauli $Z$ gate respectively.


\subsection{Shamir's $(t,n)$ threshold Secret Sharing Scheme}
Shamir's $(t, n)$ threshold secret sharing scheme~\cite{Shamir} for a dealer $D$ and $n$ participants $\{P_1, P_2, \cdots, P_n\}$, consists of following two algorithms:\\
\\
\begin{itemize}
\item{\it Share Distribution Algorithm}:
\begin{enumerate}
\item The dealer $D$ chooses a prime $d$ such that $n\leq d\leq 2n$ and his secret $a_0\in Z_d$. He randomly chooses $(a_1, a_2, \cdots, a_{t-1})\in Z_d^{t-1}$ and sets a polynomial of degree $t-1$ as $f(x) = a_0 +a_1 x + \cdots + a_{t-1} x^{t-1}.$
\item $D$ chooses $n$ nonzero and distinct elements $x_1, x_2, \cdots , x_n \in Z_d$ and publishes all of them. He sends $f(x_i)$ to the $i$th participant $P_i$ through the private channel (for $i = 1, 2, \cdots, n$).
\end{enumerate}

\vspace*{8px}
\item{\it Secret Reconstruction Algorithm}:\\
\\
Any $t$ out of $n$ participants (suppose $\{P_1, P_2, \cdots P_t\}$) takes out their shares and calculates
{\scriptsize
\begin{equation}
a_0=\sum_{i=1}^t f(x_i)\prod_{1\leq j\leq t, j\neq i}\frac{x_j}{x_j-x_i}(mod~d)
\end{equation}}
to reconstruct the dealer's secret and shares with other participants.
\end{itemize}

\section{Revisiting The Scheme Proposed by Song {\it et al.} and Its Loophole} 
\label{sec:3}
In this section, we have revisited the $(t, n)$ threshold $d$-level quantum secret sharing (QSS) scheme proposed by Song {\it et al.}~\cite{song} and the loophole as identified by Kao and Hwang~\cite{kao}.

\subsection{Revisiting the scheme proposed by Song {\it et al.}}
In this scheme a dealer Alice wants to share a secret $a_0$ among $n$ participants $Bob_1$, $Bob_2$, $\cdots$, $Bob_n$. There are three phases in this scheme, namely, initialization phase, share distribution phase and secret reconstruction phase. Initialization and secret distribution phase exploits the idea of share distribution algorithm of Shamir's $(t,n)$ threshold secret sharing scheme. Any  set of $t$ participants can reconstruct the secret. For simplicity it is assumed that the selected qualified subset is denoted by $R = \{Bob_1, Bob_2, \cdots, Bob_t\}$. The procedure of secret reconstruction phase is as follows:
\begin{enumerate}
\item $Bob_1$ (a trusted participant) prepares a $t$ particle $d$-dimensional GHZ state of the form
    {\scriptsize
    \begin{equation}
    \ket{\phi_0} =\frac{1}{\sqrt{d}}\sum_{k=0}^{d-1}\ket{k}_1\ket{k}_2\ket{k}_3\cdots\ket{k}_t
    \end{equation}} and sends the $r$th $(r = 2, 3, \cdots, t)$ particle to the $r$th participant through the authenticate quantum channel.
    \\
    \item After receiving the particle from $Bob_1$, each $Bob_r$ calculates the shadow of his share $f(x_r)$ as
        {\scriptsize
        \begin{equation}
        s_r = f(x_r)\prod_{1\leq j \leq t, j \neq r}\frac{x_j}{x_j-x_r}mod~d,
        \end{equation}} for $r = 1, 2, \ldots, t$.
    \\
    \item Each participant $Bob_r$ performs a generalized Pauli operator $U_{0, s_r}$ on his own particle, for $r = 1, 2, \cdots, t.$ Then the system of $t$ qudits becomes:
    {\scriptsize
    \begin{eqnarray}
    \ket{\phi_1} & = & \frac{1}{\sqrt{d}}\sum_{k=0}^{d-1} U_{0, s_1}\ket{k}_1 U_{0, s_2}\ket{k}_2 U_{0, s_3}\ket{k}_3\cdots U_{0, s_t}\ket{k}_t \nonumber\\
                 & = & \frac{1}{\sqrt{d}}\sum_{k=0}^{d-1}\omega^{\left(\sum_{r=1}^t s_r\right)\cdot k}\ket{k}_1\ket{k}_2\ket{k}_3\cdots\ket{k}_t.
    \end{eqnarray}}
    \item $Bob_1$ applies $QFT^{-1}$ on his own particle. He gets dealer's secret by measuring his particle in computational basis $\{\ket{0},
\ket{1}, \cdots, \ket{d-1}\}$ and shares with other participants.
\end{enumerate}

\subsection{Loophole in the above scheme}
In the QSS scheme proposed by Song {\it et al}.~\cite{song} has claimed that at the last step of secret reconstruction phase, the dealer's secret can be recovered by applying $QFT^{-1}$ on the first particle of $\ket{\phi_1}$ and measuring it in computational basis. Kao and Hwang~\cite{kao} have shown that at the end of the secret reconstruction phase, the participants can not recover the secret. To be specific, after
applying $QFT^{-1}$ on the first particle of $\ket{\phi_1}$ the
system of $t$ qudits becomes {\scriptsize
\begin{eqnarray}
\ket{\phi_2} & = & \frac{1}{\sqrt{d}}\sum_{k=0}^{d-1}\omega^{\left(\sum_{r=1}^t s_r\right)\cdot k} QFT^{-1}(\ket{k}_1) \ket{k}_2 \ket{k}_3\ldots \ket{k}_t \nonumber\\
             & = & \frac{1}{d}\sum_{k=0}^{d-1}\omega^{\left(\sum_{r=1}^t s_r\right)\cdot k}\left(\sum_{k_1 = 0}^{d-1}\omega^{-k_1\cdot
             k}\ket{k_1}_1\right)\ket{k}_2\ket{k}_3\ldots\ket{k}_t \nonumber\\
             & = & \frac{1}{d}\sum_{k=0}^{d-1}\sum_{k_1=0}^{d-1}\omega^{\left(\left(\sum_{r=1}^t s_r\right)-k_1\right)\cdot k} \ket{k_1}_1 \ket{k}_2 \ket{k}_3\ldots
             \ket{k}_t.
\end{eqnarray}}

It clearly shows that, the particles in $\ket{\phi_2}$ are still entangled. After measuring the first particle of $\ket{\phi_2}$ in the computational basis $\{\ket{0},
\ket{1}, \cdots, \ket{d-1}\}$ it results randomly one of $\{0, 1, \cdots, d-1\}$. According to the above protocol the participants consider this as a secret. The measurement result matches with actual secret with a probability $\frac{1}{d}$. Thus, the above scheme fails to reconstruct the dealer's secret.

\section{The Proposed Scheme}
\label{sec:4}
In a $(t,n)$ threshold secret sharing scheme, the dealer, Alice shares a secret ($a_0$) among $n$ participants $\{Bob_1, Bob_2, \cdots, Bob_n\}$ in such a way that any set of $t$ participants can reconstruct the secret. The secret sharing scheme is proposed in three phases: initialization phase, share distribution phase and secret reconstruction phase.

\subsection{Initialization Phase}
\begin{enumerate}
\item According to Bertrand's postulate~\cite{Hardy}, for a given $n$, Alice can find a suitable
prime $d$ satisfying $n\leq d\leq 2n$.
\item Alice sets a finite field $F=Z_d$ and her secret $a_0\in F$.
\item Alice selects $(a_1, a_2, \cdots, a_{t-1}) \in F^{t-1}$ randomly and defines a polynomial of degree $(t-1)$ as
    {\scriptsize
    \begin{equation}
    f(x) = a_0 + a_1 x + \cdots + a_{t-1} x^{t-1}.
    \end{equation}}
\end{enumerate}

\subsection{Distribution Phase}
\begin{enumerate}
\item Alice chooses $n$ distinct and nonzero values $x_1, x_2, \cdots, x_n \in F$ and publishes all the values.
\item Alice calculates $f(x_i)$ and sends to the $i$th participant through a public channel, for $i = 1, 2, \cdots, n$.
\end{enumerate}

\subsection{Reconstruction Phase}
For simplicity, we assume that the set of reconstructor is denoted by $R = \{Bob_1, Bob_2, \cdots, Bob_t\}$. Suppose $Bob_1$ is the trusted participant who initiates the reconstruction phase.
\begin{enumerate}
\item $Bob_1$ prepares $t$ sequences $S_1, S_2, \cdots, S_t$. Each of the sequence contains $m$ decoy particles randomly selected from one of the basis among $B_1=\{\ket{i}:i={0, 1, \cdots, d-1}\}$ (computational basis) and $B_2=\{QFT(\ket{i}):i={0, 1, \cdots, d-1}\}$ (Fourier basis). Then he inserts the $i$th particle of $t$-particle $d$-dimensional entangled state of the form {\scriptsize
    \begin{equation}
    \ket{\psi} = \frac{1}{\sqrt{d}}\sum_{k = 0}^{d-1}\ket{k}_1\ket{k}_2\cdots\ket{k}_t
    \end{equation}}in to the sequence $S_i$ in a random position and sends the new sequence $S_i'$ to the $i$th participant, for $i = 2, 3, \cdots, t$.
\item After the announcement of $Bob_i$ $(i = 2, 3,\cdots, t)$ that he receives the sequence of particles $S_i^{'}$, $Bob_1$ tells the position and the measurement basis of decoy photons. Now, $Bob_i$ uses the correct basis to measure the corresponding
decoy photons and declares half of the measurement results to $Bob_1$. Then, $Bob_1$ announces the initial states of the remaining half of decoy photons. Finally, they check whether
the measurement results of decoy photons are consistent with their initial states or not. If the error rate
is greater than a predetermined threshold value, they will abort the protocol; otherwise,
they will proceed to the next step.
\item $Bob_i$ $(i = 2, 3, \cdots, t)$ discards all decoy photons from $S_i^{'}$ and be left with the $i$th particle of $\ket{\psi}$. Now, $Bob_i$ $(i = 1, 2, \cdots, t)$ calculates his shadow $s_i$ from his private share $f(x_i)$ and public values $x_1, x_2, \cdots, x_n$ by
    {\scriptsize
    \begin{equation}s_i = f(x_i)\prod_{1\leq j\leq t, j\neq i}\frac{x_j}{x_j-x_i}(mod~d).
    \end{equation}}Then, he applies generalized Pauli $X$-gate (depending on his shadow) followed by $QFT$ on his own particle and measures it in computational basis $\{\ket{0},$ $\ket{1},$ $\cdots,$ $\ket{d-1}\}$. Finally, he announces his measurement result $M_i$ to $Bob_1$ through a public classical channel.
\item Now, $Bob_1$ gets $a_0'$ by calculating
{\scriptsize
\begin{equation}
a_0' = \sum_{i=1}^{t} M_i
\end{equation}}and shares it to the other participants. If the participants do not trust each other, they request the dealer Alice to send the hash value of the secret $H(a_0)$ for a known hash function $H(\cdot)$ and they verify $H(a_0') = H(a_0).$ If this equation holds, they consider $a_0$ as Alice's secret otherwise they conclude that there is at least one dishonest participant.
\end{enumerate}

\section{Correctness of Proposed Scheme}
\label{sec:5}
In this section we will show that the shared secret can perfectly be reconstructed after the secret reconstruction phase. After having the $i$th $(i = 1, 2, \cdots, t)$ particle of $\ket{\psi}$, $Bob_i$ performs the quantum Fourier transformation (QFT) on his own particle, then the system of $t$ qudits will be of the form {\scriptsize
\begin{equation}
\ket{\psi_1} = \frac{1}{d^{\frac{t+1}{2}}}\sum_{k=0}^{d-1}\left(\sum_{l_1=0}^{d-1}\omega^{kl_1}\ket{l_1}\right)\otimes\left(\sum_{l_2=0}^{d-1}\omega^{kl_2}\ket{l_2}\right)\otimes\cdots\otimes\left(\sum_{l_t=0}^{d-1}\omega^{kl_t}\ket{l_t}\right),
\end{equation}}
where $\omega = e^{\frac{2\pi i}{d}}$. Now, $Bob_i$ calculates his shadow $s_i$ and applies $U_{s_i,0}$ on his own particle, then the system of $t$ particles will be in the form:
{\scriptsize
\begin{equation}
\ket{\psi_2} =  \frac{1}{d^{\frac{t-1}{2}}}\sum_{l_1 + l_2 + \cdots + l_t = 0}\ket{l_1+s_1}\ket{l_2+s_2}\cdots\ket{l_t+s_t}.
\end{equation}}


Now, each participants $Bob_i$, measures his particle in the computational basis $\{\ket{0}, \ket{1}, \cdots,$ $\ket{d-1}\}$ and gets the measurement result as $M_i = (l_i+s_i)$ such that $l_1 + l_2 + \cdots + l_t = 0.$ $Bob_i$ ($i = 2, 3, \cdots, t$) sends their measurement result $M_i$ to $Bob_1$. Now, $Bob_1$ calculates $(M_1 + M_2 + \cdots + M_t)$ and get Alice's secret $a_0$, as
{\scriptsize
\begin{eqnarray}
\sum_{i=0}^{t}M_i & = & (l_1+s_1) + (l_2+s_2) + \cdots + (l_t+s_t)\nonumber\\
                & = & s_1 + s_2 + \cdots + s_t\nonumber\\
                & = & \left(f(x_1)\prod_{1\leq j\leq t, j\neq 1}\frac{x_j}{x_j-x_1}\right) + \left(f(x_2)\prod_{1\leq j\leq t, j\neq 2}\frac{x_j}{x_j-x_2}\right) + \cdots + \left(f(x_t)\prod_{1\leq j\leq t, j\neq t}\frac{x_j}{x_j-x_t}\right)\nonumber\\
                & = & a_0.
\end{eqnarray}}
Thus, by performing the steps described in secret reconstruction phase any set of $t$ participants among $n$ can recover the dealer's secret.

\section{Security Analysis}
\label{sec:6}
In this section, the security of proposed $(t, n)$ threshold $d$ level secret sharing scheme is analyzed. We will show that, the proposed protocol can defend both outsider and participant's attack.

\subsection{Outside Attack}
\vspace*{8px}
$\bullet$ {\it Intercept and Resend Attack}\\
\\
Suppose an eavesdropper Eve intercepts the particle sent by $Bob_1$ to $Bob_i$ $(i = 2, 3, \cdots, t)$ and resends a sequence of forged particles to $Bob_i$ to gain some information. Each particle of the GHZ state is inserted in a sequence of decoy particles which are randomly chosen from computational basis and Fourier basis. Now, an outside eavesdropper Eve does not know the position and measurement bases of decoy particles. So he will choose the basis randomly to measure the decoy particle. It will introduce an error with a probability $\frac{d-1}{2d}$ for each decoy photon. In the scheme, we have $m$ decoy photons in each sequence. Thus, the eavesdropper Eve will be detected with a probability $1-(\frac{d+1}{2d})^m$. Which will be close to 1 for large $m$.\\
\\
\noindent$\bullet$ {\it Entangle and Measure Attack}\\
\\
In this kind of attack, the attacker, Eve uses a unitary operation $U_E$ to entangle an ancillary
particle with the transmitted quantum state and then measures the ancillary particle to
steal secret information. Let us assume that the ancillary particles prepared by Eve are $E = (\ket{E_1}, \ket{E_2}, \cdots)$ and the effects of the unitary operation $U_E$ performed on the decoy
particles are shown as follows:
{\scriptsize
\begin{equation}
U_E\ket{j}\ket{E_i} = \sum_{k=0}^{d-1}a_{jk}\ket{k}\ket{e_{jk}},
\end{equation}} where $\sum_{k=0}^{d-1}|a_{jk}|^2 = 1$ and the $d^2$ states $\{\ket{e_{jk}} : j,k = 0, 1, \cdots, d-1\}$ are determined by the unitary operation $U_E$, for $j = 0, 1, \cdots, d-1$. In order to avoid the eavesdropping check, Eve
has to set $a_{jk} = 0$, for $j\neq k$ if the decoy particles are in the computational basis $B_1=\{\ket{i}:i={0, 1, \cdots, d-1}\}$. Let us denote $\ket{j'} = QFT(\ket{j})$. Now, {\scriptsize
\begin{equation}
U_E\ket{j'}\ket{E_i} = \sum_{p=0}^{d-1}\left(\sum_{k=0}^{d-1}\omega^{(j-p)k}a_{kk}\ket{e_{kk}}\right)\ket{p'}.
\end{equation}} When the decoy particles are in the Fourier basis $B_2=\{QFT(\ket{i}):i={0, 1, \cdots, d-1}\} =\{\ket{i'}:i={0, 1, \cdots, d-1}\}$, Eve has to set {\scriptsize
\begin{equation}
\sum_{k=0}^{d-1}\omega^{(j-p)k}a_{kk}\ket{e_{kk}} = 0,
\end{equation}} for $p, j = 0, 1, \cdots, d-1$ and $p\neq j$. Now, it is a system of homogeneous equations with $d$ variables $\{a_{kk}\ket{e_kk} : k = 0, 1, \cdots d-1\}$. Solving the system, we have
{\scriptsize
\begin{equation}
a_{00}\ket{e_{00}} = a_{11}\ket{e_{11}} = \cdots = a_{(d-1), (d-1)}\ket{e_{(d-1), (d-1).}}
\end{equation}}
Hence, Eve cannot distinguish between {\scriptsize $$a_{00}\ket{e_{00}}, a_{11}\ket{e_{11}}, \cdots, a_{(d-1),(d-1)}\ket{e_{(d-1),(d-1)}}$$}
and thus cannot get useful information by measuring the ancillary particles. So, the entangle-and-measure attack is defended by the scheme successfully.\\
\\
$\bullet$ {\it Man in Middle Attack}\\
\\
In this type of attack, Eve intercepts or destroys some particles and sends some forged particles, but the legitimate parties think that they are communicating directly. As Eve does not know the position of decoy particles, security analysis of the scheme against this attack is very similar to the intercept and resend attack. To resist this type of attack one can also introduce quantum identity authentication which authenticate a legitimate party and provide an outside attacker to impersonate a participant to communicate with others.\\
\\
\noindent$\bullet$ {\it Trojan Horse Attack}\\
\\
In the proposed scheme, photons are used to transmit the information. There are two types of trojan horse attack: invisible photon attack and delay photon attack. To defend invisible photon attack, the participants should add a filter before their devices to allow only the photon signals whose wavelengths are close to the operating
one to come in. In order to prevent the delay photon attack, the participants
randomly select a subset of the received photon signals as sample signals and split
each sampling signal with a photon number splitter (PNS) and measure the two signals with the base
$B_1=\{\ket{i}:i\in{0, 1, \cdots, d-1}\}$ and $B_2=\{QFT(\ket{i}):i\in{0, 1, \cdots, d-1}\}$ randomly. If the multi-photon rate is unreasonably high, the
transmission should be terminated and be repeated again from the beginning.

\subsection{Participant's Attack}
In a multiparty scheme, it is also possible to have an attack from the participants. A participants' attack is generally more powerful than an outsider attack and thus it needs more attention. In 2007, Gao {\it et al}.~\cite{gao} first analyzed the participants' attack for a multiparty quantum cryptographic protocol. Here, we will discuss the participants' attack from a single dishonest participant and the colluding attack from two or more
dishonest participants.\\
\\
$\bullet$ {\it The Participant Attack from a Single dishonest participant}\\
\\
To perform an attack, a dishonest participant $Bob_j$ (for $j\in \{2, 3, \cdots n\}$) intercepts the particle sequence $S_{j'}$, which is sent to $Bob_{j'}$, ($j\in \{2, 3, \cdots t\}$ and $j' \neq j$) from $Bob_1$. As he does not know the position and measurement basis of decoy particles he will be caught as an outsider eavesdropper. If he listens to the measurement result $M_{j'}$, then also he can not get $Bob_{j'}$'s shadow $s_{j'}$ (or private share $f(x_{j'})$) as he does not know the value of $l_{j'}$. So, $Bob_j$ (for $j\in \{2, 3, \cdots n\}$) can not get the secret alone or gain any information about other's private share.

A dishonest participant $Bob_j$, for $j\in \{1, 2, \cdots t\}$ can measure his particle in computational basis after receiving the particles from $Bob_1$, to steal some private information of other participants or to get the secret alone. Then the system of $t$ qudits will be of the form $\ket{i}_1\ket{i}_2\cdots \ket{i}_t$, where $i\in\{0, 1, \cdots, d-1\}$, i.e., if $Bob_j$ gets the state $\ket{i}$ after measurement, then he knows that the state of all other participants is also $\ket{i}$. Now, after applying $U_{s_r, 0}QFT$ by each participant $Bob_r$ $(r \in\{1, 2, \cdots, t\})$, the system of $t$ qudits will be of the form
{\scriptsize
\begin{equation}
\frac{1}{d^{\frac{t}{2}}}\sum_{l_1, l_2, \cdots, l_t = 0}^{d-1}\omega^{j(l_1+l_2+ \cdots +l_t)}\ket{l_1+s_1}\ket{l_2+s_2}\cdots\ket{l_t+s_t}
\end{equation}}
Then $Bob_r$ gets the measurement result $M_r = l_r + s_r$ (for $r = 1, 2, \cdots, t$), where $l_1, l_2, \cdots, l_t \in \{0, 1, \cdots, d-1\}$. Now, $Bob_1$ will calculates
{\scriptsize
\begin{equation}
a_0' = \sum_{r=0}^{t} M_r = \sum_{r=0}^{t}(l_r + s_r) = a_0 + \sum_{r=0}^{t}l_r.
\end{equation}}
It clearly shows that, in general $a_0 \neq a_0'$ and $a_0'$ matches with $a_0$ only when $\sum_{r=0}^{t}l_r = 0$, which occurs with a probability $d^{-(\frac{t-2}{2})}$. So, due to the participant's attack, the recovered secret is different from dealer's secret. But, the attack can be detected at the last step of the scheme when honest participants verify $H(a_0) = H(a_0')$.
%

It is also possible that, after recovering the secret, $Bob_1$ reads the secret and sends a forged value to other participants. Then also the participants can detect the eavesdropping from the hash value condition.\\
\\
$\bullet$ {\it The colluding attack from $(t-1)$ or less dishonest participants}\\
\\
In the proposed scheme, Alice considers a $(t-1)$ degree polynomial $f(x)$ and her secret is $a_0 = f(0)$. To calculate $f(0)$ for a secret polynomial $f(x)$, it is required to have a knowledge about the value of $f(x)$ for $t$ nonzero and distinct values of $x$. The dealer shares $n$ values of $f(x)$ for $n$ distinct and nonzero values of $x$ with the participants in private. So, a set of $(t-1)$ or less participants can not obtain dealer's secret's secret. It is also impossible to gain personal information of a participant by colluding two or more participants.

\section{Comparison}
\label{sec:7}
\begin{center}
\begin{table}[hb]
\caption{Comparison of QSS schemes based on QFT}
\vspace*{8pt}
{\small
\begin{tabular}{| c  c  c  c |}
\hline   & Yang {\it et al.}~\cite{yang3} & Xiao and Gao~\cite{xiao}  & Our Scheme\\
\hline $(t,n)$ or $(n,n)$ threshold & $(n,n)$ & $(n,n)$ & $(t,n)$\\
\hline No. of particles in GHZ state & $n$ & $n + 1$ & $t$\\
\hline No. of $QFT$ applied & $n$ & $n + 1$ & $t$\\
\hline No. of unitary operations & $n$ & 1 & $t$\\
\hline No. of measurement operations & $n$ & $n + 1$ & $t$\\
\hline Use of hash function & no & no & yes\\
\hline Use of decoy particles & no & yes & yes\\
\hline
\end{tabular}}
\end{table}
\end{center}
Yang {\it et al.}~\cite{yang3} has introduced the quantum Fourier transformation in quantum secret sharing. They have proposed an $(n, n)$ threshold secret sharing scheme that shares a classical secret in higher dimension. After having the shares of dealer's secret, the participants share an $n$ particle $d$-dimensional GHZ state. Then the $r$th $(r = 1, 2, \cdots, n)$ participant performs quantum Fourier transformation, a unitary operation as generalized Pauli operator $U_{0, s_r}$ and single particle measurement on his own particle respectively. Finally the participants gets the dealer's secret by calculating the sum of their measurement results. In their protocol secret distribution phase is not described.

Xiao and Gao~\cite{xiao} has proposed a $(n, n)$ threshold $d$-level quantum secret sharing scheme. This scheme uses local operations to share the secret and only classical communication to reconstruct it. At first the dealer prepares $d$ dimensional $(n+1)$-particle GHZ state. He keeps the first particle with him and shares the other $n$ particles with $n$ participants. Then the dealer applies the unitary transformation $U_{0, \beta}$ on his own particle, where $\beta$ is the dealer's secret. After that, all the participants and dealer applies quantum Fourier transformation on their own particle and measures their own particles in computational basis. After knowing the dealer's measurement result, the participants will recover the secret by adding their measurement results along with dealer's.

The above two proposals of QSS using $QFT$ are with $(n, n)$ structure. As $(t, n)$ threshold QSS schemes are more flexible than $(n, n)$ ones, Song {\it et al.}~\cite{song} has proposed a QSS scheme that shares $d$ dimensional classical secret. Note that the output of the secret reconstruction phase for the scheme described in~\cite{song}, differs from the dealer's secret with a high probability~\cite{kao}. Thus, we have proposed $(t, n)$ threshold $d$-level QSS based on $QFT$ to overcome this loophole. We have compared the performance of our protocol with the above QSS schemes using $QFT$ in the Table 1.

So our proposed $(t, n)$ threshold $d$-level scheme is more flexible, universal and practical than other QSS schemes that uses quantum Fourier transformation.

\section{Conclusion}
\label{sec:8}
In summary, we have proposed a $(t, n)$ threshold $d$-level QSS scheme based on $QFT$. The scheme uses Shamir's secret sharing scheme to distribute the shares among participants. $QFT$, generalized Pauli $Z$ gate on $d$-dimensional $n$ particle entangled state and single particle measurements on all $t$ particles are used to reconstruct the dealer's secret. This scheme can defend several outsider and participants' attack. Our scheme is more general and practical than other QSS protocols exploiting $QFT$.

\section*{Acknowledgements}
One of the author (SR) acknowledges the support from the institute in the form of institute research fellowship (Grant no: IIT/Acad/PGS$\&$R/F.II/2/15/MA/90J03) of Indian Institute of Technology Kharagpur.




\end{document}